\documentclass[conference]{IEEEtran}
\IEEEoverridecommandlockouts
\usepackage{cite}
\usepackage{amsmath,amssymb,amsfonts}
\usepackage{siunitx}
\usepackage{algorithm,algorithmic}
\usepackage{graphicx}
\usepackage{textcomp}
\usepackage{xcolor}
\usepackage{hyperref}
\usepackage{tabularx, booktabs}
\usepackage{marvosym}
\usepackage{cleveref}
\usepackage{pifont, lipsum}
\usepackage{stfloats}
\usepackage[table]{xcolor}
\usepackage{colortbl}
\usepackage{textcase}
\usepackage{worldflags}

\def\BibTeX{{\rm B\kern-.05em{\sc i\kern-.025em b}\kern-.08em
    T\kern-.1667em\lower.7ex\hbox{E}\kern-.125emX}}
    
\begin{document}
\title{High-Endurance UCAV Propulsion System: A 1-D CNN-Based Real-Time Fault Classification for Tactical-Grade IPMSM Drive}
\author{
	\IEEEauthorblockN{Tahmin Mahmud}
	\IEEEauthorblockA{
		Elmore Family School of Electrical and Computer Engineering, Purdue University, West Lafayette, IN 47907, USA \\
		Email: mahmud13@purdue.edu
	}
}

\maketitle

\begin{abstract} High-performance propulsion for mission-critical applications demands unprecedented reliability and real-time fault resilience. Conventional diagnostic methods (signal-based analysis and standard ML models) are essential for stator/rotor fault detection but suffer from high latency and poor generalization across variable speeds. This paper proposes a 1-D Convolutional Neural Network (CNN) framework for real-time fault classification in the HPDM-350 interior permanent magnet synchronous motor (IPMSM). The proposed architecture extracts discriminative features directly from high-frequency current and speed signals, enabling sub-millisecond inference on embedded controllers. Compared to state-of-the-art long short term memory (LSTM) and classical ML approaches, the 1-D CNN achieves a superior weighted $F_1$-score of 0.9834. Validated through high-fidelity magnetic-domain MATLAB/Simscape models, the method demonstrates robust performance across a \(\pm\) 2700 RPM envelope, providing a lightweight solution for mission-critical electric propulsion systems.

\end{abstract}

\begin{table*}[!b] 
\centering
\caption{State-of-the-Art MALE UCAV Propulsion: Conventional Platforms vs. Next-Gen MQ-9 Reaper}
\label{tab1}
\setlength{\tabcolsep}{2pt}
 \begin{tabular}{||l|c|c|c|>{\columncolor{green!35}}c||} 
 \hline
 \textbf{Feature} & \textbf{MQ-9 Reaper} \cite{b7} & \textbf{Rainbow CH-5} \cite{b8} & \textbf{Orion-E} \cite{b9} & \textbf{Next-Gen MQ-9 Reaper} \\
\hline\hline
Origin & USA \raisebox{-0.15cm}{\resizebox{!}{0.4cm}{\worldflag{US}}} & China \raisebox{-0.15cm}{\resizebox{!}{0.4cm}{\worldflag{CN}}} & Russia \raisebox{-0.15cm}{\resizebox{!}{0.4cm}{\worldflag{RU}}} & USA  \raisebox{-0.15cm}{\resizebox{!}{0.4cm}{\worldflag{US}}} \\
\hline
Machine Type & Honeywell TPE331-10 Turboprop & WJ-9A/AEP50E Turboprop & RU APD-110 Turbo-piston & HPDM-350 IPMSM \\
\hline
Fuel Type & Std. Mil Jet Fuel (JP-8) & Diesel or Jet Fuel (RP-3/RP-5) & Diesel or Jet Fuel & $H_{\mathrm{2}}$ fuel cell tanks and/or SSBPs \\
\hline
Drive System & Mechanical Gearbox & Mechanical Gearbox & Mechanical Gearbox & Direct Drive (gearless) \\
\hline
Peak Power & $\sim$700 kW (940 hp) & $\sim$450--500 kW & $\sim$85--100 kW & 350 kW (scalable to 700+ kW) \\
\hline
Inverter & N/A (mechanical) & N/A (mechanical) & N/A (mechanical) & Integrated (SiC-based) \\
\hline
Power Density & $\sim$1--2 kW/kg (system level) & $<$1.5 kW/kg & $<$1 kW/kg & 7 kW/kg (continuous) \\
\hline
Cooling & Oil/Air (large radiators) & Liquid/Air & Liquid/Air & Internal Liquid Cold Plate \\
\hline
Fault Tolerance & Single Engine (no redundancy) & Single Engine & Single Engine & Multisector (dual/quad redundant) \\
\hline
\end{tabular}
\end{table*}

\section{Introduction} \label{sec1}
 The Department of War (DOW) has prioritized the development of next-generation autonomous platforms capable of operating in contested environments where stealth capabilities and acoustic signature management are paramount \cite{b1}. The key to achieving these tactical advantages is the reliability of the propulsion system. However, the transition toward high-power-density electrification introduces significant vulnerabilities. Faults within the main traction machine and its associated drive electronics are catastrophic, potentially leading to mission failure or loss of high-value assets. While permanent magnet synchronous motors (PMSMs) offer superior power density \cite{b2, b3}, they are susceptible to critical anomalies such as stator inter-turn short circuits (ITSC), demagnetization of the rotor magnets (DMAG), and inverter-stage open-circuit faults (IOC). As a result, real-time fault diagnostics and failure modes are critical to ensure the readiness and survival of tactical unmanned combat aerial vehicle (UCAV) fleets.

\begin{figure}[!htbp]
	\centering
	\includegraphics[width=\columnwidth]{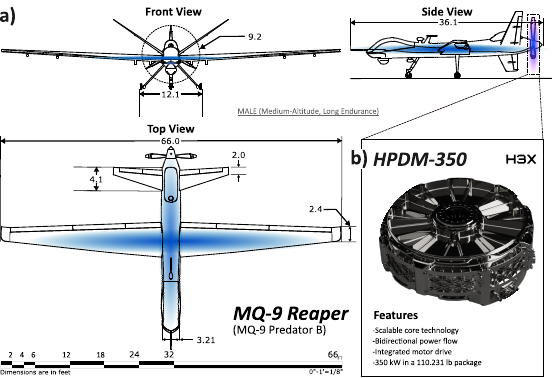} 
	\caption{MQ-9 Reaper Powertrain. (a) Three-view dimensional layout of the atomic airframe, (b) Close-up $2\times$ stacked HPDM-350 IDS.}
	\label{fig1}
\end{figure}

The current MQ-9 Reaper (Predator B) \cite{b4} architecture as shown in Fig.~\ref{fig1}(a) relies on a legacy turboprop configuration primarily the Honeywell TPE331-10GD \cite{b5}. This engine presents limitations in terms of thermal management, mechanical complexity, and high acoustic signature. By leveraging the HPDM-350 interior-PMSM (IPMSM) developed by H3X Technologies \cite{b6}, the platform undergoes a fundamental shift toward a more robust, high-performance electrified and integrated drive system (IDS) architecture. Unlike legacy systems, the HPDM-350 combines the motor and power electronics into a single, scalable package as shown in Fig.~\ref{fig1}(b), significantly enhancing the power-to-weight ratio and providing the bidirectional power flow necessary for advanced energy management. This transition not only improves the platform's endurance but also necessitates the implementation of sophisticated, real-time fault classification to maintain the integrity of the tactical-grade drive system. Table~\ref{tab1} presents a state-of-the-art comparison of medium-altitude long endurance (MALE) UCAV propulsion systems, highlighting the features of the Next-Gen MQ-9 Reaper.

\begin{figure*}[htbp]
	\centering
	\includegraphics[width=\textwidth, keepaspectratio]{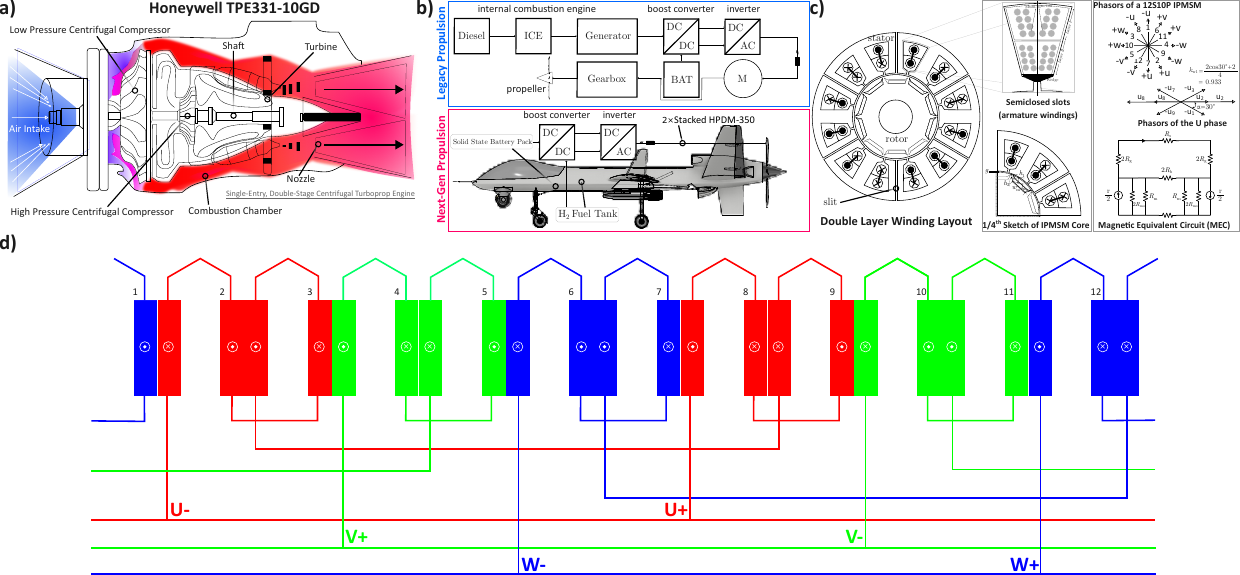}
	\caption{System-level integration of the proposed UCAV propulsion. (a) Honeywell TPE331-10GD turboprop engine anatomy, (b) Propulsion system evolution: legacy vs next-gen, (c) HPDM-350's conceptual machine cross-section and phasor analysis, (d) Double-layer unfolded winding layout for the 12S stator.}
	\label{fig2}
\end{figure*}

Recent PMSM fault diagnosis has shifted toward data-driven frameworks to achieve real-time reliability and robustness against parameter drift in nonlinear environments. To bypass traditional model inaccuracies, Wang et al. \cite{b10} utilize a hybrid LSTM-Transformer model for high-fidelity drive dynamics prediction. Similarly, the authors in \cite{b11} employ a purely data-driven CNN for flywheel PMSM fault classification using raw current signatures. Addressing multi-sensor integration, Wang et al. \cite{b12} propose a non-invasive framework using Multi-signal Gramian Angular Difference Fields (MGADF) and a Dung Beetle Optimized Random Forest to resolve noise sensitivity in distinguishing similar faults like demagnetization and eccentricity. For scenarios with scarce labeled data, the paper in \cite{b13} introduces a semi-supervised Distance-Preserving Self-Organizing Map (DP-SOM) to map high-dimensional health indicators onto low-dimensional manifolds. Further addressing data scarcity in aviation, Shih et al. \cite{b14} implement a few-shot learning (FSL) metric-based classifier to identify overlapping mixed faults with minimal samples. To bridge the gap between physical constraints and data, the authors in \cite{b15} combines physics-based modeling with data-driven indicators for actuator degradation analysis, while \cite{b16} utilizes statistically interpretable spectral methods for anomaly detection. The authors in \cite{b17}  provides a standardized, large-scale experimental dataset to facilitate consistent benchmarking of these emerging machine-learning classifiers. The reviewed literature shows a shift from model-based to hybrid learning for PMSM fault diagnosis, highlighting the need for a 1-D CNN architecture enabling real-time, accurate fault detection in UCAV IPMSMs. The critical contributions of this paper are as follows:

\begin{enumerate}
	\item All-electric $H_{\mathrm{2}}$ fuel cell UCAV as the 
      target deployment platform for the proposed system-level fault classifier.
    \item Validated performance using a 12S10P magnetic domain specifically parameterized for the 350~kW HPDM-350.
    \item Achieved 22.9$\times$ and 7.7$\times$ reduced inference time over SVM and random forest (RF) baselines respectively.
\end{enumerate}

Subsequent sections are organized as follows: Section \ref{sec2} reviews the foundational concepts, Section \ref{sec3} presents the proposed fault classification framework, Section \ref{sec4} conducts a case study, Section \ref{sec5} presents post-training accuracy and comparative performance benchmarks. Finally, Section \ref{sec6} concludes the paper and outlines future research directions.

\vspace{-0.18cm}
\section{Preliminaries} \label{sec2}
This section establishes the technical foundation for the proposed cloud-based diagnostic framework by analyzing the limitations of current turboprop systems and the architectural advantages of next-generation electrified drives.

\subsection{Honeywell TPE331-10GD Turboprop and its Challenges}
The Honeywell TPE331-10GD is a legacy propulsion system widely used in tactical UCAVs, featuring a single-shaft centrifugal compressor and axial turbine as illustrated in Fig.~\ref{fig2}(a). Although it offers a high power-to-weight ratio, its gearbox complexity and continuous combustion produce high thermal and acoustic signatures, reducing stealth. Limited modularity also restricts advanced flight control and energy management integration.

\subsection{From Legacy ICE to Next-Gen All-Electric Propulsion}
UCAV propulsion is shifting from ICE-based systems with mechanical gearboxes to high-power-density electric drives. Conventional designs use diesel generators and battery-buffered DC buses. The proposed architecture replaces this with hydrogen ($H_{\mathrm{2}}$) fuel cells and a solid-state battery pack powering dual HPDM-350 IPMSMs via a boost-integrated converter as presented in Fig.~\ref{fig2}(b), improving efficiency and redundancy while introducing new power-electronic failure modes requiring real-time monitoring.

\subsection{HPDM-350 IPMSM Modeling}

Based on the datasheet \cite{b18}, the HPDM-350 is assumed to be a 12-slot, 10-pole (12S10P) IPMSM designed for high torque density using a double-layer fractional-slot concentrated winding (FSCW) layout. Due to limited public disclosure of the HPDM-350 core architecture, the machine is approximated as an IPMSM with a double-layer distributed winding for analytical and simulation studies. To accurately classify faults, the electromagnetic (EM) behavior is modeled using a magnetic equivalent circuit (MEC) that accounts for the flux linkage $\lambda_{\mathrm{dq}}$ and the cross-coupling effects between the $d$- and $q$-axes. The voltage equations in the synchronous 
reference frame are defined as:

\begin{equation} 
    u_{\mathrm{d}} = R_{\mathrm{s}} i_{\mathrm{d}} 
    + L_{\mathrm{d}} \frac{d i_{\mathrm{d}}}{dt} 
    - \omega_{\mathrm{e}} L_{\mathrm{q}} i_{\mathrm{q}}
    \label{eq1}
\end{equation}

\begin{equation}
    u_{\mathrm{q}} = R_{\mathrm{s}} i_{\mathrm{q}} 
    + L_{\mathrm{q}} \frac{d i_{\mathrm{q}}}{dt} 
    + \omega_{\mathrm{e}} \left( L_{\mathrm{d}} i_{\mathrm{d}} + \psi_{\mathrm{f}} \right)
    \label{eq2}
\end{equation}

\noindent where $R_\mathrm{s}$ is the stator resistance, $\omega_{\mathrm{e}}$ is the electrical angular velocity, and $\psi_{\mathrm{f}}$ is the permanent magnet flux linkage. For the 12S10P double-layer configuration, the winding factor $k_\mathrm{w1}$ is optimized to suppress space harmonics, and the mutual inductance $M$ between phases is significantly reduced due to the concentrated nature of the coils. The phase-belt distribution for the U-phase, as illustrated in the phasor diagram of Fig.~\ref{fig2}(c), follows a specific spatial sequence represented by the winding function $N(\theta)$, where the magnetomotive force (MMF) distribution is:

\begin{equation}
    F(\theta, t) = 
    \sum_{\substack{n=1,3,5,\ldots}}^{\infty} 
    \frac{4}{\pi n} 
    N_{\mathrm{ph}}\, k_{\mathrm{w}n} 
    \sin(n\theta)\, i(t)
    \label{eq3}
\end{equation}

\noindent where $N_{\mathrm{ph}}$ is the number of turns per phase, $k_{\mathrm{w}n}$ is the $n$-th order winding factor, and $\theta$ is the spatial angle along the stator bore. We present the double-layer unfolded winding layout for the 12S stator in Fig.~\ref{fig2}(d). This high-fidelity modeling of the winding layout is critical for the 1-D CNN framework to distinguish between healthy operation and subtle distortions caused by inter-turn short circuits or demagnetization faults.

\subsection{Machine Fault Types and Mathematical Characterization}

Fault diagnosis in the HPDM-350 requires a precise understanding of how each 
fault condition perturbs the nominal EM behavior. A fault signal 
comparison and FFT spectrum per fault class (Phase~A, Tooth~1, RPM~=~1500) is shown in Fig.~\ref{fig3}(a). In this section, we characterize four critical fault types relevant to the machine under test (MUT). 

\subsubsection{Demagnetization Fault}
DMAG occurs when the permanent magnet flux density falls below its rated 
value due to thermal stress, overcurrent, or component ageing. The degraded flux linkage 
$\psi_{\mathrm{f}}^{\mathrm{dem}}$ is modeled as,

\begin{equation}
  \psi_{\mathrm{f}}^{\mathrm{dem}} = (1 - \mu_{\mathrm{d}})\, \psi_{\mathrm{f}}
    \label{eq4}
\end{equation}

\noindent where $\mu_{\mathrm{d}} \in [0,1)$ is the demagnetization severity factor. The resulting torque deficit is expressed as,

\begin{equation}
    T_{\mathrm{e}}^{\mathrm{dem}}  = \frac{3}{2} \times \mathrm{p} \left[ \psi_{\mathrm{f}}^{\mathrm{dem}}\, i_\mathrm{q} + 
    \left( L_\mathrm{d} - L_\mathrm{q} \right) i_\mathrm{d}\, i_\mathrm{q} \right]
    \label{eq5}
\end{equation}

\noindent where $\mathrm{p}$ is the number of pole pairs. As $\mu_{\mathrm{d}}$ increases, the back-EMF 
amplitude decreases proportionally, producing a measurable reduction in the 
fundamental harmonic of the phase current spectrum.

\subsubsection{Ground Fault (Coil-to-Ground Short)}
A ground fault (GND) occurs when the insulation between a stator winding and the machine 
frame breaks down, creating a low-impedance path to ground. For a fault in phase-$\mathrm{a}$ 
with fault resistance $R_\mathrm{g}$, the modified phase voltage equation becomes,

\begin{equation}
    u_\mathrm{a} = R_\mathrm{s} i_\mathrm{a} + L_\mathrm{s} \frac{di_\mathrm{a}}{dt} + e_\mathrm{a} - R_\mathrm{g}\, i_\mathrm{g}
    \label{eq6}
\end{equation}

\noindent where $i_\mathrm{g}$ is the GND current and $e_\mathrm{a}$ is the back-EMF of phase-$\mathrm{a}$. The resulting current unbalance index $\delta_\mathrm{I}$ across the three phases is expressed as,

\begin{equation}
    \delta_\mathrm{I} = 
    \frac{\max\left(|I_\mathrm{a}|, |I_\mathrm{b}|, |I_\mathrm{c}|\right) - I_\mathrm{avg}}
    {I_\mathrm{avg}} \times 100\%
    \label{eq7}
\end{equation}

\noindent where $I_\mathrm{avg}$ is the mean of the three-phase (3P) current magnitudes. A 
significant rise in $ \delta_\mathrm{I}$ serves as a primary diagnostic indicator for 
coil-to-ground faults.

\subsubsection{Isolated Turn Fault (Inter-Turn Short Circuit)}
An inter-turn short circuit (ITSC) arises when the insulation between adjacent turns 
within the same coil degrades, forming a shorted loop of $\mu_\mathrm{t}$ turns out of a total $N_\mathrm{ph}$ turns per phase. The faulty phase resistance and inductance are modified as,

\begin{equation}
    R_{\mathrm{s}}^{\mathrm{f}} = R_\mathrm{s} \left(1 - \frac{\mu_\mathrm{t}}{N_\mathrm{ph}}\right)^2 + R_\mathrm{sc}
    \label{eq8}
\end{equation}

\begin{equation}
    L_{\mathrm{s}}^{\mathrm{f}}  = L_\mathrm{s} \left(1 - \frac{\mu_\mathrm{t}}{N_\mathrm{ph}}\right)^2
    \label{eq9}
\end{equation}

\noindent where $R_\mathrm{sc}$ is the resistance of the shorted loop. The circulating 
short-circuit current $i_\mathrm{sc}$ within the faulted turns is,

\begin{equation}
    i_\mathrm{sc} = \frac{\mu_\mathrm{t}}{N_\mathrm{ph}} \cdot \frac{e_\mathrm{a}}{R_\mathrm{sc} + j\omega_\mathrm{e} L_\mathrm{sc}}
    \label{eq10}
\end{equation}

\noindent This localized circulating current produces concentrated ohmic heating 
$P_\mathrm{sc} = i_\mathrm{sc}^2 R_\mathrm{sc}$, which accelerates insulation degradation and is detectable as a sideband harmonic at $(1 \pm 2k)f_\mathrm{s}$ in the current spectrum, where $f_\mathrm{s}$ is the fundamental supply frequency and $k = 1, 2, 3, \ldots$ $\infty$

\subsubsection{Open-Circuit Fault}
An open circuit fault (IOC) occurs when a phase terminal or inverter switch disconnects, 
reducing the effective number of active phases. For a 3P IPMSM with phase-$\mathrm{c}$ open-circuited ($i_\mathrm{c} = 0$), the constraint imposed on the remaining phases is,

\begin{equation}
    i_\mathrm{a} + i_\mathrm{b} = 0 \implies i_\mathrm{b} = -i_\mathrm{a}
    \label{eq11}
\end{equation}

The EM torque under this condition degrades to:
\begin{equation}
    T_{\mathrm{e}}^{\mathrm{oc}} = \mathrm{p} \times \, \psi_\mathrm{f} \left( i_\mathrm{a} \sin\theta_\mathrm{e} - i_\mathrm{b} \sin\left(\theta_\mathrm{e} 
    - \frac{2\pi}{3}\right) \right)
    \label{eq12}
\end{equation}

\noindent where $\theta_\mathrm{e}$ is the electrical rotor position. The loss of the third 
phase introduces a second-order torque ripple at $2\omega_\mathrm{e}$, which can be represented as an elevated vibration and thermal stress in the remaining active windings, and is identifiable through asymmetry in the normalized current space vector trajectory.

\section{DL Pipeline: 1-D CNN Fault Classifier} 
 \label{sec3}

This section presents the proposed end-to-end deep-learning (DL) pipeline
for the EM fault classification that encompasses physics-informed dataset generation,
frame-based signal preprocessing, a lightweight one-dimensional
convolutional neural network~(1-D~CNN), and supervised training.
\subsection{Dataset Preparation}

\subsubsection{HPDM-350 IPMSM Operating Principle}

The MUT is a 12S10P IPMSM, configured at the slot level to preserve the magnetic asymmetry introduced by each fault class. The motor is parameterized as presented in Table~\ref{tab2}. The motor model presents an open-loop generator (dynamometer configuration) at constant shaft speed, which permits controlled data collection across a wide speed range without field-oriented control (FOC)
dynamics masking the fault signatures.

\begin{table}[h!]
\centering
\caption{Design Space: HPDM-350 IPMSM}
\label{tab2}
\footnotesize
\setlength{\tabcolsep}{4pt} 
\begin{tabular}{||l | l||}
\hline
\textbf{Parameter} & \textbf{Value} \\
\hline\hline
Motor type & IPMSM \\
Pole pairs, $\mathrm{p}$ & 5 \\
Maximum torque, $T_{\mathrm{max}}$ & 1238 Nm \\
Maximum power, $P_{\mathrm{max}}$ & 350 kW \\
Stator resistance, $R_\mathrm{s}$ & 7 m$\Omega$ \\
d-axis inductance, $L_\mathrm{d}$ & 0.20 mH \\
q-axis inductance, $L_\mathrm{q}$ & 0.35 mH \\
Average inductance, $L_\mathrm{s}$ & 0.25 mH \\
Permanent magnet flux, $\mathrm{PM}$ & 0.50 Wb \\
Time step, $dt$ & 1 $\mu$s \\
Speed control step, $dt_{\mathrm{speed}}$ & 100 $\mu$s \\
Electrical cycle & 72$^\circ$ \\
Current vector, $I_{\mathrm{vec}}$ & [0, 100, 200] A \\
Angle vector, $\theta_{\mathrm{vec}}$ & 0–72$^\circ$ \\
Beta vector, $B_{\mathrm{vec}}$ & -180°:90°:180° \\
Iron loss coefficients & $k_\mathrm{hs}=1.2\times10^{-3}$ \\
 & $k_\mathrm{Js}=0.8\times10^{-5}$ \\
 & $k_\mathrm{es}=0.5\times10^{-6}$ \\
\hline
\end{tabular}
\end{table}

The mechanical speed is swept uniformly from
$-\SI{2700}{rpm}$ to $+\SI{2700}{rpm}$ in steps of
$\SI{150}{rpm}$, yielding 37~operating points that span both motoring
and regenerative quadrants.
For each~(fault, speed) pair, $N_\mathrm{s} = 50{,}000$ samples are generated at
an effective output rate of $f_\mathrm{s} = \SI{50}{\kilo\hertz}$ (base
time-step $\Delta t = \SI{1}{\micro\second}$).
The electrical frequency at any speed $n$~(r.p.m.) is expressed as,
\begin{equation}
  f_\mathrm{e} = \frac{N_\mathrm{p} \, n}{60}
  \label{eq:fel}
\end{equation}
ranging from \SI{0}{\hertz} to \SI{225}{\hertz} at the extremes of the
sweep.

\subsubsection{Fault Class Definitions}

The HPDM-350 IPMSM fault scenarios include healthy operation with balanced 3P currents; DMAG reducing PM flux by 35\%, lowering torque and introducing $2^\mathrm{nd}$-order THDs; phase-$\mathrm{a}$ GND, causing DC offset and $3^\mathrm{rd}$-order perturbations; ITSC affecting 5\% of phase-$\mathrm{a}$, generating $5^\mathrm{th}$-order THD currents; and IOC phase-$\mathrm{a}$, halving torque and redistributing current with second-harmonic ripple. All simulated waveforms include additive white Gaussian noise~(AWGN) at a signal-to-noise ratio (SNR)  of \SI{40}{\decibel} to reflect realistic
inverter switching and sensor noise conditions.

\vspace{0.5cm}

\subsubsection{Simulated Signal Set}

Each scenario produces seven time-series channels, arranged as the transpose of a row vector,
{\footnotesize
\begin{equation}
  \mathbf{s}(t) =
  \bigl[\,\tau(t),\; i_{\mathrm{a}}(t),\; i_{\mathrm{b}}(t),\; i_{\mathrm{c}}(t),\;
         v_{\mathrm{a}}(t),\; v_{\mathrm{b}}(t),\; v_{\mathrm{c}}(t)\,\bigr]^{\!\top}
  \!\in \mathbb{R}^{7}
  \label{eq:channels}
\end{equation}
}
where $\tau$ is the instantaneous EM torque,
$i_\mathrm{a,b,c}$ are the 3P stator currents, and $v_\mathrm{a,b,c}$ are
the 3P terminal voltages. These seven signals are written row-wise to a comma-separated values in a \texttt{.csv} file with the schema:

\begin{center}
\small
\texttt{rpm, fault, torque, ia, ib, ic, va, vb, vc}
\end{center}

\noindent The dataset contains a total of $37 \times 5 \times 50{,}000 = 9{,}250{,}000$ rows. Simulated signals of the MUT produced in the MATLAB/Simscape environment are exported as a \texttt{.csv} file which is loaded locally in the Jupyter notebook via Pandas as:

{\scriptsize
\begin{verbatim}
import pandas as pd
CSV_PATH = 'HPDM350_fault_data.csv'

df = pd.read_csv(CSV_PATH)
print(f"Loaded: {df.shape[0]:,} rows × {df.shape[1]} columns")
print(f"Faults: {df['fault'].value_counts().to_dict()}")
print(f"RPM range: {df['rpm'].min()} to {df['rpm'].max()}")
df.describe().round(3)
\end{verbatim}
}

\subsubsection{Train/Val Speed Split}

The dataset comprises $11{,}470$ frames uniformly distributed across five fault classes and 37~operating speeds spanning $\pm$2700~rpm, ensuring balanced representation for unbiased classifier training. The dataset is partitioned into a training subset (every alternate value: $\pm 150,\pm 300,\ldots,\pm 2700$~rpm, 25~values in total) and a held-out test subset
(intermediate values: $\pm 75,\pm 225,\ldots,\pm 2625$~rpm,
17~values in total). The network therefore never observes the test speeds during optimization.
\subsection{Frame-Based Data Preprocessing}
\label{subsec:preprocessing}

Raw waveforms are segmented into fixed-length overlapping frames
to form independent training or inference instances.

\subsubsection{Sliding-Window Segmentation}

For a raw signal $\mathbf{S} \in \mathbb{R}^{N_\mathrm{s} \times 7}$ sampled at $f_\mathrm{s}$, frames
$\mathbf{F}_k \in \mathbb{R}^{T_\mathrm{f} \times 7}$ are obtained as
\begin{equation}
  \mathbf{F}_k = \mathbf{S}[k \cdot \Delta : k \cdot \Delta + T_\mathrm{f}, \; :], 
  \quad k = 0,1,2,\dots
  \label{eq:frame}
\end{equation}
with $T_\mathrm{f} = 1000$ samples, hop $\Delta = T_\mathrm{f} - T_\mathrm{ov} = 800$ samples ($T_\mathrm{ov} = 200$, 20\% overlap). At $f_\mathrm{s} = \SI{50}{kHz}$, each frame spans \SI{20}{ms}, covering multiple electrical cycles even at low speeds. Overlap ensures faults near frame edges are captured.

\begin{figure*}[htbp]
	\centering
	\includegraphics[width=1.04\textwidth, keepaspectratio]{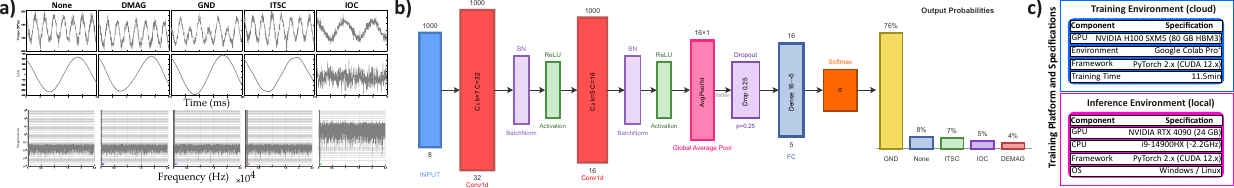}
	\caption{Signal-to-decision fault diagnostic chain. (a)~Fault signatures \& frequency spectra, 
(b)~\textbf{FaultCNN1D} architecture, and (c)~Hardware specifications.}
	\label{fig3}
\end{figure*}

\subsubsection{RPM Channel Augmentation}

Normalized shaft speed $\hat{n} = n / n_\mathrm{max}$ ($n_\mathrm{max} = \SI{2700}{rpm}$) is appended as an eighth channel:
\begin{equation}
  \tilde{\mathbf{F}}_k =
  \bigl[\,\mathbf{F}_k \;\big|\; \hat{n}\,\mathbf{1}_{T_\mathrm{f}}\,\bigr]
  \in \mathbb{R}^{T_\mathrm{f} \times 8}
  \label{eq:rpm_channel}
\end{equation}
enabling the CNN to decouple speed-dependent harmonics from fault signatures.

\subsubsection{Per-Channel Standardization}

Each channel is standardized using training statistics:
\begin{equation}
  \hat{x}_{c} = \frac{x_{c} - \mu_c^{\mathrm{train}}}{\sigma_c^{\mathrm{train}} + \epsilon}, 
  \quad c = 1, \ldots, 8
  \label{eq:norm}
\end{equation}
with $\epsilon = 10^{-8}$. Validation and test frames use the same statistics. Frames are finally permuted to channels-first layout $\tilde{\mathbf{F}}_k \in \mathbb{R}^{8 \times T_\mathrm{f}}$ for \texttt{torch.nn.Conv1d}.

\subsection{1-D CNN Framework}
\label{subsec:architecture}

\subsubsection{Design Rationale}

A lightweight 1-D CNN framework as illustrated in Fig.~\ref{fig3}(b) is used for fault classification because motor signals are temporal, with diagnostic information in waveform \emph{shape} and \emph{frequency} rather than magnitude. The 1-D convolutional kernel slides along each channel, learning local temporal features (e.g., harmonic patterns, current spikes, torque dips) that are translation-equivariant. Global average pooling collapses the time dimension, producing a fixed-length, translation-\emph{invariant} descriptor directly separable by the classifier. This compact architecture generalizes well, making it ideal for embedded motor drive units.

\subsubsection{Layer Configuration}

The proposed \textbf{FaultCNN1D} has two convolutional blocks, a GAP layer, dropout, and a linear classification head. Input $\mathbf{X} \in \mathbb{R}^{C_\mathrm{in} \times T}$ with $C_\mathrm{in}=8$ and $T=1000$.

\paragraph{Block 1}
\texttt{Conv1d} with $C_1=32$, $k_1=7$, $p_1=3$ extracts fine-grained harmonic features. Output $\mathbf{A}_1 \in \mathbb{R}^{32 \times 1000}$ passes through \texttt{BatchNorm1d} and \texttt{ReLU}.

\paragraph{Block 2}
\texttt{Conv1d} with $C_2=16$, $k_2=5$, $p_2=2$ abstracts higher-level temporal patterns. Output $\mathbf{A}_2 \in \mathbb{R}^{16 \times 1000}$ again passes through \texttt{BatchNorm1d} and \texttt{ReLU}.

\paragraph{\text{Global Average Pooling}}
\texttt{AdaptiveAvgPool1d(1)} computes
\begin{equation}
  z_j = \frac{1}{T} \sum_{t=1}^{T} a_{2,j}(t), \quad j=1,\ldots,16,
  \label{eq:gap}
\end{equation}
yielding $\mathbf{z} \in \mathbb{R}^{16}$.

\paragraph{Dropout and Classification}
\texttt{Dropout} with $p=0.25$ regularizes the network. A fully connected layer maps $\mathbf{z}$ to $\hat{\mathbf{y}} \in \mathbb{R}^5$, converted to posterior probabilities \cite{b19} via
\begin{equation}
  P(\mathcal{F}_c \mid \tilde{\mathbf{F}}_k)
  = \frac{\exp(\hat{y}_c)}{\sum_{j=1}^{5} \exp(\hat{y}_j)},
  \quad c \in \text{\scriptsize $\{\textsf{None}, \textsf{DEMAG}, \textsf{GND}, \textsf{ITSC}, \textsf{IOC}\}$}
  \label{eq:softmax}
\end{equation}

\subsubsection{Parameter Count}

Learnable parameters are constrained for real-time inference:
\begin{align}
  \Theta_{\mathrm{Conv1}} &= C_\mathrm{in} C_1 k_1 + 2C_1 = 8 \cdot 32 \cdot 7 + 64 = 1{,}856, \notag\\
  \Theta_{\mathrm{Conv2}} &= C_1 C_2 k_2 + 2C_2 = 32 \cdot 16 \cdot 5 + 32 = 2{,}592, \notag\\
  \Theta_{\mathrm{FC}}    &= C_2 N_\mathrm{cls} + N_\mathrm{cls} = 16 \cdot 5 + 5 = 85, \notag\\
  \Theta_{\mathrm{total}} &= 1{,}856 + 2{,}592 + 85 = 4{,}533.
  \label{eq:params}
\end{align}
Including batch-norm scalars ($4(C_1+C_2)=192$) gives 4{,}725 parameters, requiring \SI{18.5}{KB} at FP32.

\subsubsection{Architecture and Training Hyperparameter Summary}

Table~\ref{tab3} summarizes the complete network topology and
all training hyperparameters used in this work.

\begin{table}[h!]
\centering
\caption{FaultCNN1D Architecture and Training Hyperparameters}
\label{tab3}
\resizebox{\columnwidth}{!}{%
\begin{tabular}{||l | c | c||}
\hline
\textbf{Parameters} & \textbf{Configuration} & \textbf{Value / Output} \\
\hline\hline
\multicolumn{3}{||c||}{\textbf{Network Architecture}} \\
\hline
Input & Sequence & $8 \times 1000$ \\
Block 1 & Conv1d ($k=7$, $C=32$) + BN + ReLU & $32 \times 1000$ \\
Block 2 & Conv1d ($k=5$, $C=16$) + BN + ReLU & $16 \times 1000$ \\
GAP & AdaptiveAvgPool1d(1) & $16 \times 1$ \\
Dropout & $p_{\mathrm{drop}}$ & 0.25 \\
FC & Linear & $16 \rightarrow 5$ \\
Output & Softmax & 5 classes \\
Total parameters & & 4{,}725 \\
Model size (FP32) & & 18.5 KB \\
\hline
\multicolumn{3}{||c||}{\textbf{Preprocessing Parameters}} \\
\hline
Frame length & $T_f$ & 1{,}000 samples \\
Frame overlap & $T_{\mathrm{ov}}$ & 200 samples \\
Hop size & $\Delta$ & 800 samples \\
Frame duration & & 20 ms \\
Input channels & & 8 (7 signals + RPM) \\
\hline
\multicolumn{3}{||c||}{\textbf{Training Hyperparameters}} \\
\hline
Optimizer & & AdamW \\
Learning rate & $\eta_0$ & $1 \times 10^{-3}$ \\
LR schedule & CosineAnnealingLR & $\eta_{\min} = 10^{-6}$ \\
Weight decay & $\lambda$ & $1 \times 10^{-4}$ \\
Label smoothing & $\varepsilon_{\mathrm{ls}}$ & 0.05 \\
Batch size & & 256 \\
Max epochs & & 10{,}000 \\
Validation frequency & & Every 500 iterations \\
Train/Val split & & 80\% / 20\% \\
Train RPMs & & 25 speed points \\
Test RPMs & & 12 unseen speed points \\
\hline
\end{tabular}%
}
\end{table}

\subsection{Model Training}
\label{subsec:training}

\subsubsection{Loss Function}
The network minimizes the cross-entropy loss with label smoothing:
\begin{equation}
  \mathcal{L}_{\mathrm{CE}}
  = -\sum_{c=1}^{C} q_c \log P(\mathcal{F}_c \mid \tilde{\mathbf{F}}_k),
  \label{eq:loss}
\end{equation}
where $q_c = (1-\varepsilon_\mathrm{ls})\,\delta_{c,y} + \varepsilon_\mathrm{ls}/C$, $\delta_{c,y}$ is the Kronecker delta, $C=5$, and $\varepsilon_\mathrm{ls}=0.05$. Label smoothing improves calibration and robustness to label noise.

\subsubsection{Optimizer (AdamW)}
Parameters are updated using Adam with Weight Decay (AdamW) optimizer \cite{b20}:
\begin{align}
  \mathbf{m}_t &= \beta_1 \mathbf{m}_{t-1} + (1-\beta_1)\,\nabla_\theta \mathcal{L}_t,
  \label{eq:adam_m}\\
  \mathbf{v}_t &= \beta_2 \mathbf{v}_{t-1} + (1-\beta_2)(\nabla_\theta \mathcal{L}_t)^2,
  \label{eq:adam_v}\\
  \theta_t     &= \theta_{t-1} - \eta_t \frac{\hat{\mathbf{m}}_t}{\sqrt{\hat{\mathbf{v}}_t}+\epsilon} - \eta_t \lambda \theta_{t-1},
  \label{eq:adamw}
\end{align}
with bias-corrected moments $\hat{\mathbf{m}}_t, \hat{\mathbf{v}}_t$, $\beta_1=0.9$, $\beta_2=0.999$, $\epsilon=10^{-8}$, $\eta_0=10^{-3}$, and $\lambda=10^{-4}$.

\subsubsection{Learning Rate Schedule}
A cosine-annealing schedule is used:
\begin{equation}
  \eta_t = \eta_{\min} + \frac{1}{2}(\eta_0-\eta_{\min}) \left[ 1 + \cos\!\left(\frac{\pi t}{T_{\max}}\right)\right],
  \label{eq:cosine_lr}
\end{equation}
with $\eta_{\min}=10^{-6}$ and $T_{\max}=10{,}000$~epochs, allowing precise convergence to flat minima.

\subsubsection{Regularisation \& Training Protocol}
To reduce overfitting, we set: (i) weight decay to $\lambda=10^{-4}$, (ii) dropout $p_{\mathrm{drop}}=0.25$ before the classification head, and (iii) gradient clipping constrains global $L_2$ norm to 1. 

The model is trained for $10{,}000$~epochs with a mini-batch size of 256. Validation accuracy is evaluated every 500~optimizer iterations over a randomly selected 20\,\% hold-out partition of the training-speed data (stratified by class). The checkpoint corresponding to the highest validation accuracy is retained as the final model. A detailed hardware deployment specifications is shown in Fig.~\ref{fig3}(c).

\section{Case Study: HPDM 350 IPMSM Steady-Sate Operation \& Dynamic Response} \label{sec4}
To  validate the proposed magnetic domain implementation of the MUT, the dynamic performance of the digital twin (DT) \cite{b21} FEM-parameterized HPDM-350 IPMSM driven via a high-performance FOC modulation scheme is evaluated under high-fidelity simulation environments in MATLAB/Simscape.

\begin{figure*}[!t]
	\centering
	\includegraphics[width=\textwidth]{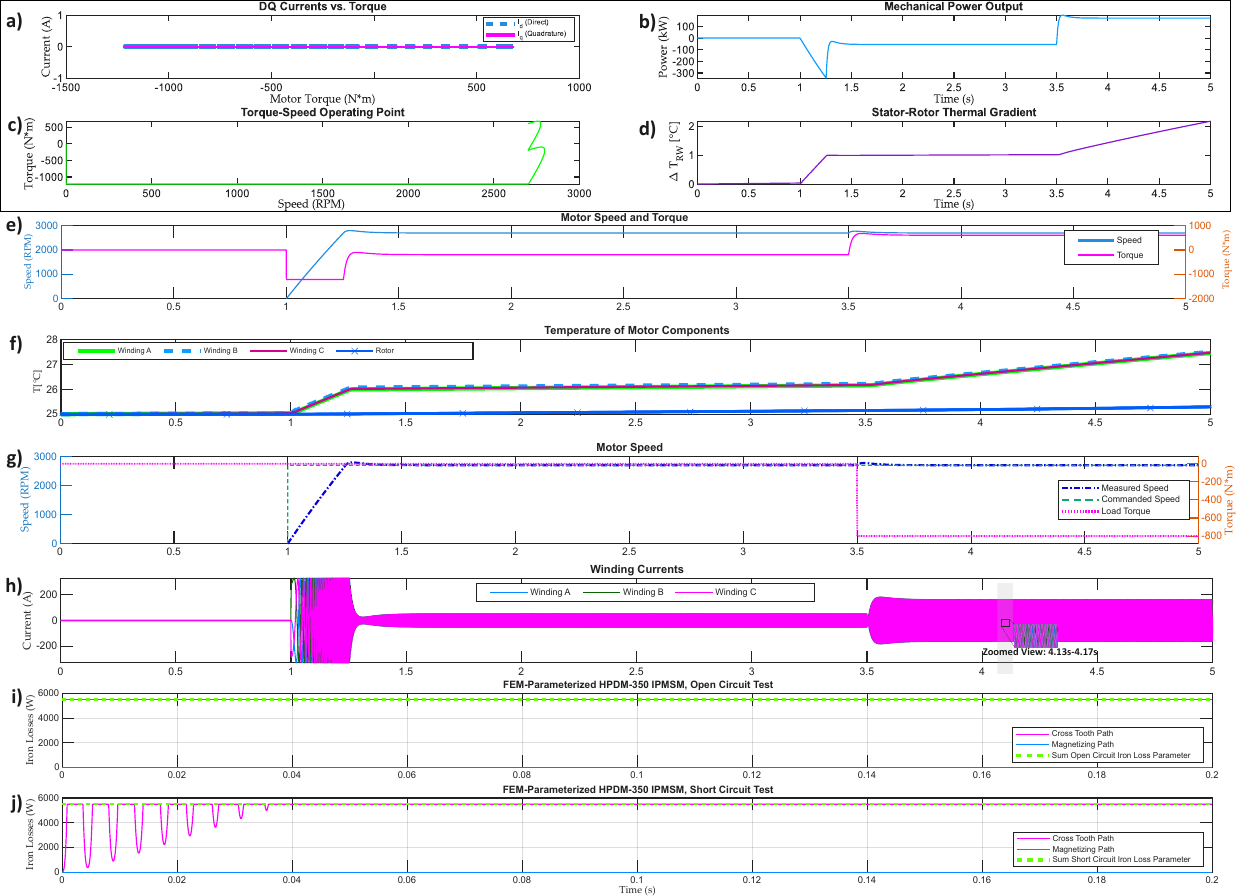}
	\caption{Dynamic performance and thermal characteristics of the HPDM-350 IPMSM. (a) DQ current vs. EM torque, (b) Output power profile, (c) Torque-speed envelope, (d) Thermal gradient $\Delta T_\mathrm{RW}$, (e) Speed and torque response, (f) Temp. distribution, (g) Speed profile, (h) 3P currents, (i) IOC \& (j) ITSC-core loss.}
	\label{fig4}

	\vspace{0.42cm} 
	\includegraphics[width=\textwidth]{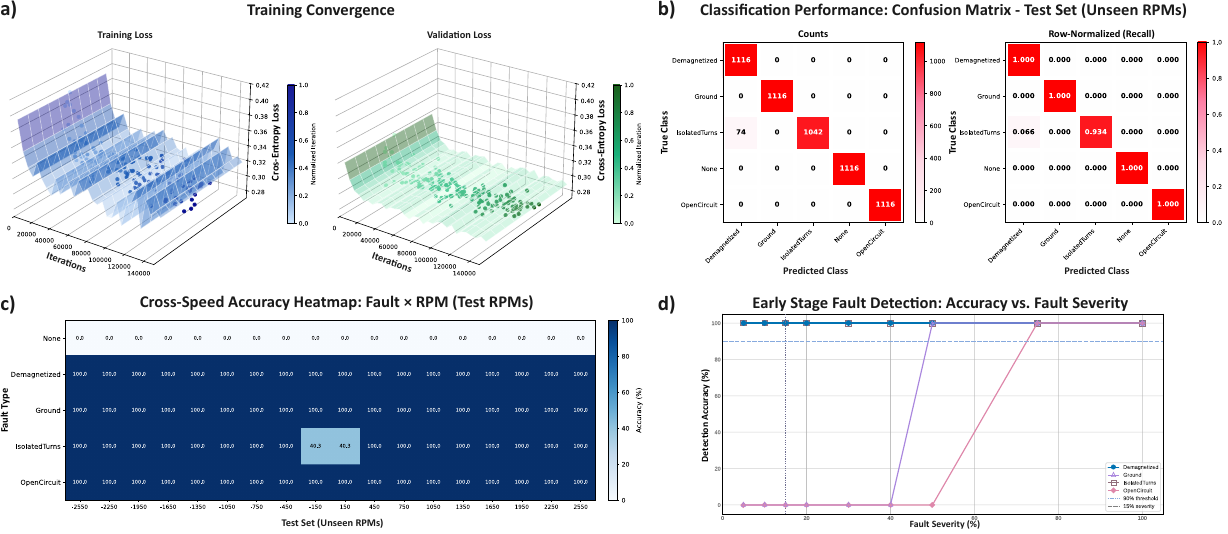}
	\caption{Post-training performance evaluation. (a)~Training and Validation convergence surfaces, (b)~Confusion matrices (counts and row-normalized recall), (c)~Cross-speed accuracy heatmap, and (d)~Incipient-fault detection accuracy as a function of injected fault severity.}
	\label{fig5}
\end{figure*}

Fig.~\ref{fig4}(a) illustrates the relationship between the DQ-axis currents and the generated electromagnetic torque, where the overlapping trajectories at the origin indicate that the controller effectively manages the $i_\mathrm{d}$ and $i_\mathrm{q}$ components to maintain zero torque during no-load intervals. 

The mechanical power output profile presented in Fig.~\ref{fig4}(b) illustrates the high-performance transition of the HPDM-350 between motoring and generating quadrants. During the startup acceleration phase between 1.0s and 1.3s, the system draws significant electrical energy to produce the required acceleration torque to reach the +2700~RPM target. A sharp power dip to $-$350~kW is observed at approximately 1.3s, representing a transient period of regenerative braking as the controller extracts kinetic energy to prevent speed overshoot, effectively operating in the second quadrant where speed and torque are in opposite directions. Following this transient, the system enters a steady-state load rejection phase from 3.5s onwards, where the application of an $-$800~Nm load triggers an increase in torque production to maintain the commanded speed. This results in a sustained jump to +180~kW, characterizing standard first-quadrant motoring operation where the HPDM-350 performs work on the load. This bidirectional energy flow validates the robustness of the control logic under both dynamic acceleration and sudden load perturbations.

The operational envelope is detailed in Fig.~\ref{fig4}(c), showing the torque-speed operating point transitioning toward the high-speed region at +2700 RPM, while Fig.~\ref{fig4}(g) confirms precise speed tracking with minimal overshoot despite load torque perturbations. Thermal stability is critical for the HPDM-350 due to its ultra-compact 50~kg mass. Fig.~\ref{fig4}(d) and Fig.~\ref{fig4}(f) monitor the thermal gradient and component temperatures, respectively. The results show a gradual rise in temperature for the windings and rotor, stabilizing near 27.5$^\circ$C under the tested cooling flow rate, validating the liquid-cooling effectiveness modeled in the MATLAB/Simscape thermal network. Transient current behavior is presented in Fig.~\ref{fig4}(h), where the phase currents show a significant expansion during the high-torque interval, followed by a stabilized steady-state region. The zoomed-in view reveals balanced sinusoidal waveforms, indicating that the 12S magnetic domain model accurately captures slot-level harmonics. Finally, the iron loss analysis in Fig.~\ref{fig4}(i) and Fig.~\ref{fig4}(j) compares the IOC and ITSC core-loss distributions. The short-circuit test shows significantly higher periodic iron loss peaks, exceeding 5000~W which emphasizes the increased magnetic flux density and core saturation occurring within the cross-tooth paths during fault conditions. This comprehensive analysis demonstrates that the DT IPMSM model maintains high power density and thermal integrity across its +2700~RPM operating range.

\section{Post-Training Accuracy and Performance}
\label{sec5}

Beyond conventional held-out test-set evaluation, the trained
\textbf{FaultCNN1D} model is subjected to a wide range of validation
experiments to assess its practical readiness for deployment. 

\textbf{Training Convergence:}
Fig.~\ref{fig5}(a) presents the 3D cross-entropy loss
surfaces for both the training and validation sets plotted against the
normalized iteration axis.
Both surfaces descend sharply within the first 5\% of the training
budget and then plateau at comparable loss values, confirming
that the optimizer converged without exhibiting divergence or
instability. The close proximity of the two surfaces throughout the training indicates
a negligible generalization gap, providing evidence that the adopted
regularization strategy is sufficient to prevent overfitting.

\textbf{Classification Performance:}
The confusion matrices in Fig.~\ref{fig5}(b) present the model's
predictions on the held-out test set.
The raw-count matrix confirms that four of the five fault classes:
\textsf{DMAG}, \textsf{GND}, \textsf{None}, and
\textsf{IOC} are classified with zero misclassifications,
yielding per-class recall values of 1.000 as seen in the row-normalized
matrix.
The \textsf{ITSC} class exhibits a recall of 0.934, with 74
frames misclassified as \textsf{DMAG}.
This confusion is physically interpretable: both faults induce
asymmetric flux perturbations in phase-$\mathrm{a}$ that produce overlapping
harmonic content at low fault severity, approaching the noise floor
of the 35~dB SNR operating condition.

\textbf{Cross-Speed Generalization:}
Fig.~\ref{fig5}(c) displays the per-class accuracy evaluated
independently at each of the 17~unseen test-speed points spanning
$-$2700 to $+$2700~rpm.
With the exception of two intermediate speed points at which
\textsf{ITSC} accuracy falls to 40.3\,\%, all remaining
fault-speed combinations achieve 100\,\% classification accuracy. This confirms that that the proposed frame-based conditioning approach, specifically the normalized RPM channel appended as an eighth input, 
enabling effective decoupling of speed-dependent harmonic content from
fault-specific signatures across the full operating envelope.

\textbf{Incipient-Fault Detection:}
Fig.~\ref{fig5}(d) illustrates the detection accuracy as the injected
fault amplitude is progressively scaled from 5\% to 100\% of its
rated magnitude.
\textsf{DMAG}, \textsf{GND}, and \textsf{None} achieve the
90\% detection threshold at or below 15\% fault severity, while
\textsf{IOC} requires approximately 40\% severity and
\textsf{ITSC} requires approximately 50\% severity before
reliable detection is achieved.
These results confirm that the model is capable of incipient-fault
diagnosis for the majority of fault classes, with the detection
threshold directly governed by the SNR of the
fault-induced harmonic relative to the ambient noise floor. A training summary along with
quantitative accuracy, $F_1$-score, parameter count, and inference
latency comparisons against SVM, Random Forest, KNN, and LSTM
baselines are presented in Table~\ref{tab4}.

\begin{table}[h!]
\centering
\caption{Comparative Performance and Training Summary}
\label{tab4}
\resizebox{\columnwidth}{!}{%
\begin{tabular}{||l | c | c | c | c||}
\hline
\textbf{Method / Metric} & \textbf{Accuracy (\%)} & \textbf{F1-Score} & \textbf{Params} & \textbf{Inf. Time (ms)} \\
\hline\hline
\multicolumn{5}{||c||}{\textbf{Comparative Performance Benchmarks}} \\
\hline
\textbf{1-D CNN (Ours)} & \textbf{98.34} & \textbf{0.9834} & \textbf{4,533} & \textbf{0.00079} \\
Random Forest & 96.39 & 0.9636 & N/A & 0.00607 \\
SVM (RBF) & 94.44 & 0.9446 & N/A & 0.01808 \\
KNN ($k=7$) & 89.56 & 0.8980 & N/A & 0.00820 \\
LSTM (2-layer) & 84.97 & 0.8568 & 13,989 & 0.00093 \\
\hline
\multicolumn{5}{||c||}{\textbf{CNN Training and Generalization Summary}} \\
\hline
\textbf{Parameters} & \multicolumn{2}{c|}{\textbf{Configuration}} & \multicolumn{2}{c||}{\textbf{Value / Output}} \\
\hline
Final Training Accuracy & \multicolumn{2}{c|}{$\mathrm{Acc_{train}}$} & \multicolumn{2}{c||}{96.07\%} \\
Final Validation Accuracy & \multicolumn{2}{c|}{$\mathrm{Acc_{val}}$} & \multicolumn{2}{c||}{95.86\%} \\
Generalization Gap & \multicolumn{2}{c|}{$\Delta_{\mathrm{gen}}$} & \multicolumn{2}{c||}{0.21\% (Low)} \\
Best Validation Accuracy & \multicolumn{2}{c|}{$\mathrm{Acc_{best}}$} & \multicolumn{2}{c||}{96.39\%} \\
Total Training Duration & \multicolumn{2}{c|}{$t_{\mathrm{total}}$} & \multicolumn{2}{c||}{11.5 min} \\
\hline
\end{tabular}%
}
\end{table}

\section{Conclusion \& Future Work} \label{sec6}
This study presents a real-time fault diagnosis framework for the HPDM-350 IPMSM using a lightweight 1-D CNN architecture. By integrating slot-level magnetic domain modeling with deep learning, the system achieves a weighted $F_1$-score of 0.9834 in detecting complex winding and rotor faults with an ultra-low inference time of 0.79 $\mu$s. The 0.21\,\% generalization gap between training and validation
accuracy confirms that the model does not overfit to the training-speed distribution. These results validate the feasibility of deploying deep-learning diagnostics on resource-constrained embedded platforms for SiC-driven motor drives. Future work includes experimental testing and hardware-in-the-loop (HIL) validation using physical HPDM-350 prototype to evaluate reliability under real-world noise and multiple-fault profiles.

\end{document}